\documentclass[superscriptaddress,amsmath,amssymb,aps,prl]{revtex4-2}

\usepackage{graphicx}
\usepackage{dcolumn}
\usepackage{bm}
\usepackage{hyperref}
\usepackage{notes2bib}
\usepackage{siunitx}
\usepackage[usenames,dvipsnames]{xcolor}

\setcitestyle{super}

\newcommand{\nj}{\ensuremath n_\mathrm{j}}
\newcommand{\nleads}{\ensuremath n_\mathrm{\ell}}
\newcommand{\twist}{\ensuremath \theta=1.06^\circ}
\newcommand{\Ic}{\ensuremath I_\mathrm{c}}
\newcommand{\Is}{\ensuremath I_\mathrm{s}}
\newcommand{\Ir}{\ensuremath I_\mathrm{r}}

\newcommand{\ns}{\ensuremath n_\mathrm{s}}
\newcommand{\Bp}{\ensuremath B_\perp}
\newcommand{\Lj}{\ensuremath L_\mathrm{j}}
\newcommand{\vf}{\ensuremath v_\mathrm{F}}

\begin{document}

\title{Gate-Defined Josephson Junctions in Magic-Angle Twisted Bilayer Graphene}

\author{Folkert K. de Vries}
\email{devriesf@phys.ethz.ch}
\author{El\'ias Portolés}	
\author{Giulia Zheng}
\affiliation{Solid State Physics Laboratory, ETH Zurich,~CH-8093~Zurich, Switzerland}
\author{Takashi Taniguchi}
\affiliation{International Center for Materials Nanoarchitectonics, National Institute for Materials Science,  1-1 Namiki, Tsukuba 305-0044, Japan}
\author{Kenji Wantanabe}
\affiliation{Research Center for Functional Materials, National Institute for Materials Science, 1-1 Namiki, Tsukuba 305-0044, Japan}
\author{Thomas Ihn}
\author{Klaus Ensslin}
\author{Peter Rickhaus}
\email{peterri@phys.ethz.ch}
\affiliation{Solid State Physics Laboratory, ETH Zurich,~CH-8093~Zurich, Switzerland}

\maketitle

\textbf{ 
In the past two years, magic-angle twisted bilayer graphene~\cite{Suarez2010,Bistritzer2011,Nam2017,Guinea2019} has emerged as a uniquely versatile experimental platform that combines metallic, superconducting, magnetic and insulating phases in a single crystal~\cite{Cao2018,Cao2018_2,Yankowitz2018,Sharpe2019,Lu2019,nuckolls2020}. In particular the ability to tune the superconducting state with a gate voltage opened up intriguing prospects for novel device functionality. Here we present the first demonstration of a device based on the interplay between two distinct phases in adjustable regions of a single magic-angle twisted bilayer graphene crystal. We electrostatically define the superconducting and insulating regions of a Josephson junction and observe tunable DC and AC Josephson effects~\cite{Josephson1962,Shapiro1963}. We show that superconductivity is induced in different electronic bands and describe the junction behaviour in terms of these bands, taking in consideration interface effects as well.
Shapiro steps, a hallmark of the AC Josephson effect and therefore the formation of a Josephson junction, are observed.
This work is an initial step towards devices where separate gate-defined correlated states are connected in single-crystal nanostructures. We envision applications in superconducting electronics~\cite{tinkham2004,Braginski2019} and quantum information technology~\cite{Wendin2017,Liu2019} as well as in studies exploring the nature of the superconducting state in magic-angle twisted bilayer graphene.
}\newline

Gate-tunable superconductivity is a topical field of pure and applied physics research~\cite{DeSimoni2018}, recently boosted by the realization of two-dimensional (2D) superconductors~\cite{Saito2016}. 
Switching superconductivity electrostatically on and off makes device design more flexible and circumvents interface and fabrication challenges met when working with multi-material nanostructures.
Superconductivity in 2D has been realised for example at the LaAlO$_3$/SrTiO$_3$ interface~\cite{Caviglia2008}, and in van der Waals materials such as WTe$_2$~\cite{Fatemi2018,Sajadi2018}.
Because of the large charge carrier densities in most of these systems, in-situ electrostatic control is limited, and therefore gate-tunable Josephson junctions (JJs) have been realised only in a few systems~\cite{Caviglia2008,Kononov2020} where tunability is achieved by proximity to another material or by structural confinement.

Magic-angle twisted bilayer graphene (MATBG) is an ideal 2D platform for in-situ gate-tunable superconductivity~\cite{Cao2018,Yankowitz2018,Lu2019}
While there are numerous potential device applications~\cite{tinkham2004,Braginski2019,Wendin2017,Liu2019}, progress towards practical implementations has been hindered by the need for well-defined gated regions. 
Our multilayer gate technology has enabled us to create a device in a single MATBG crystal with electrostatically defined regions displaying distinct quantum phases.
This device represents a new technology that can be used to build integrated 2D electronics using correlated states~\cite{Cao2018,Cao2018_2,Yankowitz2018,Sharpe2019,Lu2019}.

\begin{figure}
\centering
\includegraphics[width=0.5\textwidth]{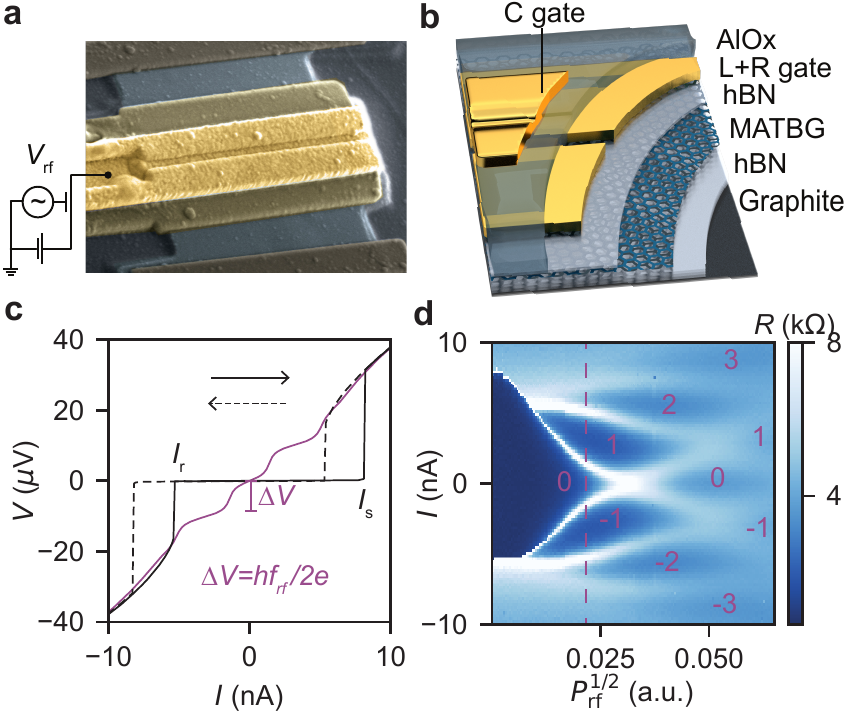}

\caption{\textbf{DC and AC Josephson effect in a MATBG gate-defined Josephson junction.}
	\textbf{a}, False-coloured, tilted scanning electron microscopy image of a similar device and  \textbf{b}, cross-section diagram of the device. A $\SI{100}{nm}$-long JJ is formed by tuning the MATBG into the superconducting state with the graphite bottom-gate and left (L) and right (R) top-gate electrodes, while the center (C) gate tunes the MATBG locally into a non-superconducting state. 
	The gate electrodes are separated from the MATBG by insulating hexagonal boron nitride (hBN), and from each other by an aluminium-oxide (AlO$\mathrm{x}$) layer.
	\textbf{c}, The applied current $I$ and measured voltage $V$ (black) reveal a switching and retrapping current ($\Is$ and $\Ir$) at density $\nj=\SI{-4.6e12}{cm^{-2}}$ in the junction area.
	Applying a voltage drive ($V_\mathrm{rf}$) at frequency $f_\mathrm{rf}$=\SI{5}{GHz} to gate C induces Shapiro steps (violet).
	\textbf{d}, differential resistance $R$ as a function of $I$ and the square root of power ($P_\mathrm{rf}^{1/2}$), showing the Shapiro step spectrum of the JJ. The step number is indicated, as is the trace plotted in \textbf{c}.
}\label{fig:1}

\end{figure}

Here we report a first step towards superconducting devices based on distinct quantum phases within a single crystal, by engineering a tunable JJ electrostatically defined in MATBG (Fig.~\ref{fig:1}a,b).
We confirm the formation of a JJ by observation of both the DC Josephson effect, with the typical hysteresis found for an underdamped JJ~\cite{tinkham2004}, and the AC Josephson effect, see Figs.~\ref{fig:1}c,d. 
For the latter we apply a radio-frequency (rf) voltage excitation ($V_\mathrm{rf}$), leading to the observation of Shapiro steps~\cite{Shapiro1963}, that are proportional to the frequency $f_\mathrm{rf}$ and the superconducting flux quantum $h/2e$. Furthermore, we demonstrate that the device can be tuned in situ from the short to the long JJ regime, establishing the flexibility of our approach. 

The device introduced here has three different gate layers that enable electrostatically defining several JJs (Fig.~\ref{fig:1}a,b). 
The global carrier density $n$ in the MATBG can be changed by a graphite bottom gate.
A pair of local top-gates (L, R) controls the density of the leads $\nleads$ (when tuned together with the bottom gate) and defines a junction of length $\Lj=\SI{100}{nm}$ and width $W=\SI{2.3}{\micro m}$. The density in this junction $\nj$ is gated by gate C.
In order to form JJs with different lengths $\Lj$ we tune another gate or a combination of gates out of the superconducting state, which changes the regions of the device corresponding to densities $\nleads$ and $\nj$ accordingly.
All measurements are performed at a temperature $T=\SI{25}{mK}$ (unless stated otherwise) in a two-terminal setup,
where we apply a current $I$ and measure the voltage $V$, from which the differential resistance $R$ is calculated. 
We correct for contact and lead resistances in the $I$--$V$ linetraces, but not in $R$ colour maps.
More details on the device fabrication and measurement setup are presented in Methods and Supplementary Information (SI), Sec.~\ref{SI:device}.

\begin{figure}
\centering
\includegraphics[width=0.5\textwidth]{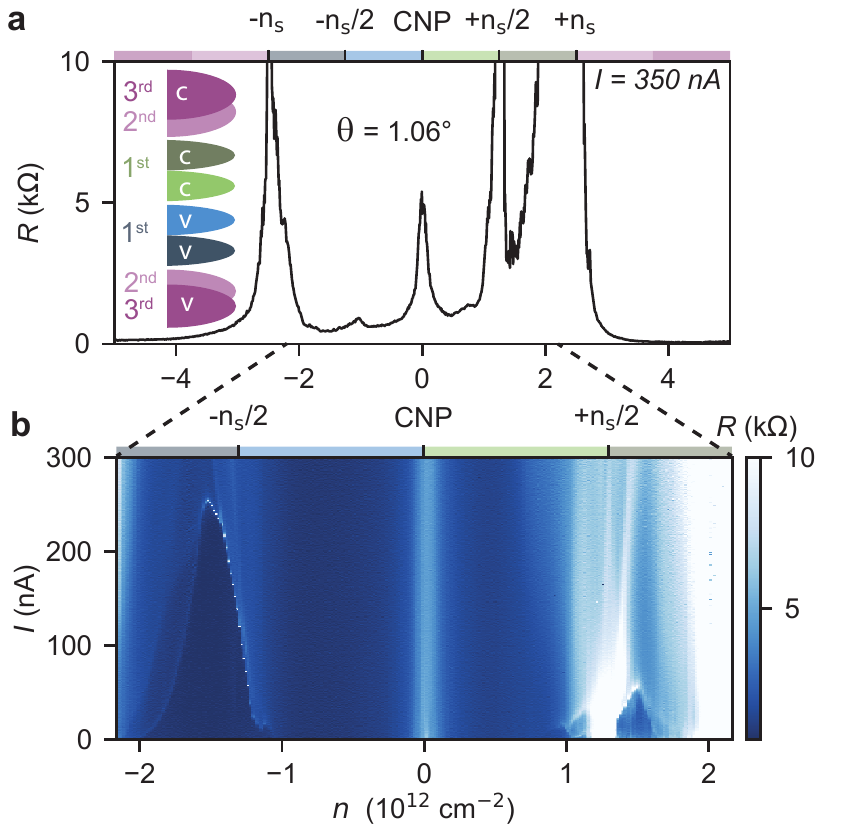}
\caption{\textbf{Bulk superconductivity in MATBG.}
	\textbf{a} $R$ as a function of $n$ shows pronounced peaks for $\pm\ns$ and $\pm\ns/2$, symmetrically around the CNP.
	The inset shows a schematic representation of the energy bands in MATBG, where the flat valence (v) and conduction (c) bands (blue and green) and higher moir\'{e} bands (violet) are indicated.
	The colours on top of the panels indicate the relation between the density and the bands.
	\textbf{b} $R$ as a function of $I$ and $n$ reveals superconductivity around $\pm\ns/2$.
}\label{fig:2}
\end{figure}

\begin{figure}
\centering
\includegraphics[width=\textwidth]{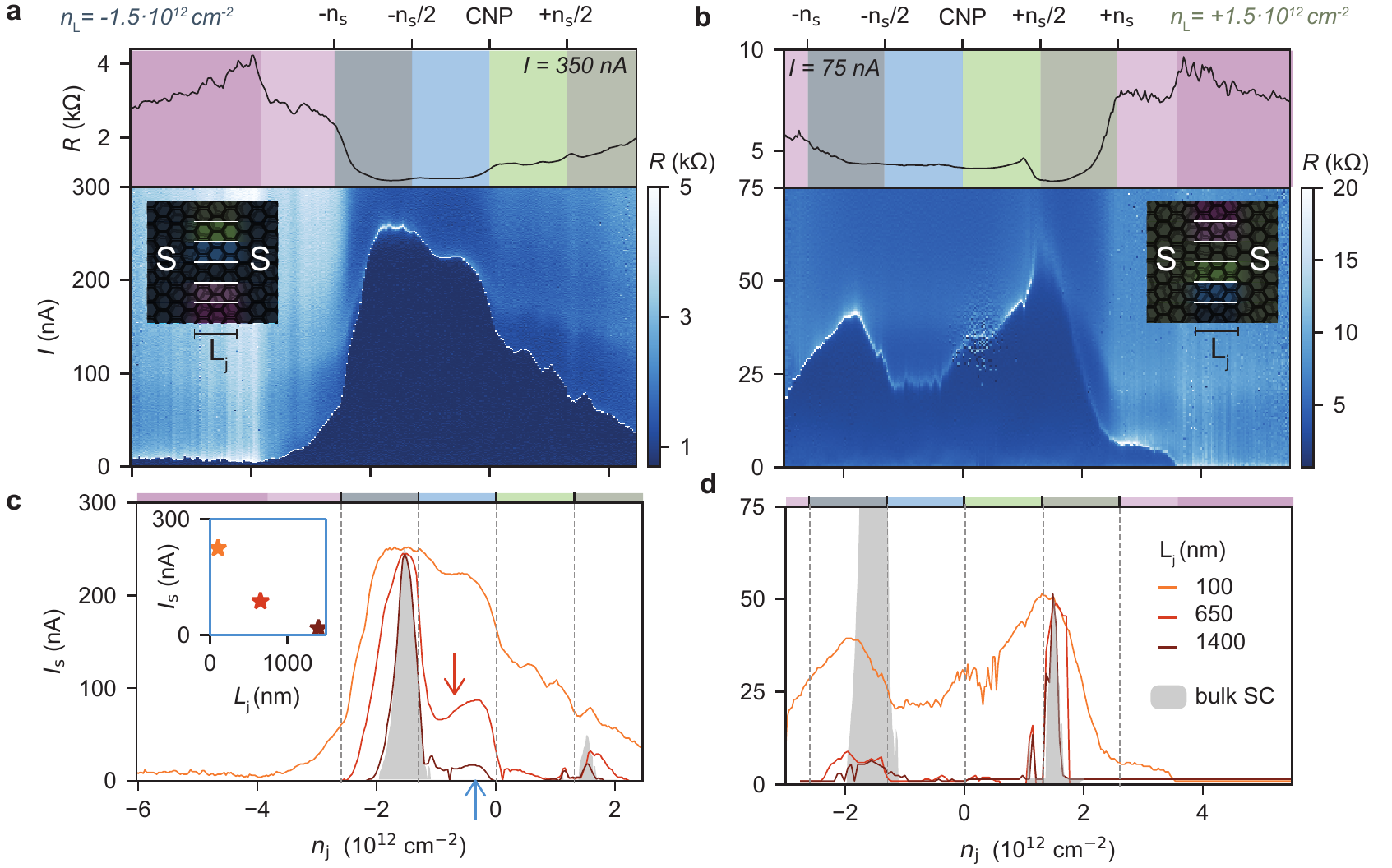}
\caption{\textbf{Gate-defined and tunable Josephson junctions.}
	\textbf{a,b}, Gate-defined JJs with $\Lj$= \SI{100}{nm} and $\nleads$ tuned to either the negatively (a) or positively doped (b) superconducting domes as indicated on the top axis. The line trace, taken at the $I$ value indicated, and colour-map plot $R$ as a function of $\nj$ (and $I$). Overlaid are the colours indicating of energy bands sketched in Fig.~\ref{fig:1}a.
	\textbf{c,d}, $\Is$ for different $\Lj$ as a function of $\nj$. 
	Bulk $\Ic$ versus $n$ is plotted for reference (grey). The inset displays $\Is$ versus $\Lj$ for $\nj$ indicated by the blue arrow.
}\label{fig:3}
\end{figure}

Before returning to the JJ we introduce the energy bands and the bulk superconducting state in our MATBG device.
In Fig.~\ref{fig:2}a we show $R$ as a function of carrier density $n$. Peaks appear in $R$ at the charge-neutrality point (CNP), and at full ($\ns$) and half-filling of the moir\'{e} bands, corresponding to a twist angle of $\twist$. 
Based on this observation we sketch the energy bands and gaps~\cite{Suarez2010,Bistritzer2011,Nam2017,Guinea2019} (inset Fig.~\ref{fig:2}a), containing the 1$^{\text{st}}$ moir\'{e} bands (blue and green), commonly referred to as the flat bands, with correlated insulator gaps at $\ns/2$~\cite{Cao2018_2,Lu2019}. 
The 1$^{\text{st}}$ band is separated from the dispersive 2$^{\text{nd}}$ and 3$^{\text{rd}}$ moir\'{e} bands (violet) by a band gap.
Note that the 2$^{\text{nd}}$ and 3$^{\text{rd}}$ band are separated in energy, which follows from measurements obtained in the JJ-device configuration (Fig~\ref{fig:3}a,b), but contradicts existing band-structure calculations~\cite{Suarez2010,Bistritzer2011,Nam2017,Guinea2019}.
To reveal the superconducting behaviour of the device we measure the differential resistance $R$ as a function of current $I$ and density $n$ (Fig.~\ref{fig:2}b). 
We observe plateaus and the typical peak in $R$ at critical supercurrent $\Ic$, where the device switches out of the bulk superconducting state. 
The largest $\Ic$ is observed at $n = \SI{-1.5e12}{cm^{-2}}$ in the superconducting dome around $-\ns/2$.
Around $+\ns/2$ a smaller superconducting dome is observed.
This in accordance with earlier reported phase diagrams for MATBG of $\twist$~\cite{Cao2018,Yankowitz2018}.
We estimate a superconducting coherence length $\xi= \SI{70}{nm}$ from $\Ic$ measurements as a function of temperature $T$ and perpendicular magnetic field $\Bp$ (SI Sec.~\ref{SI:IcTB}).

Having established superconductivity in the MATBG device, we describe the effect of detuning $\nj$ while keeping $\nleads$ superconducting.
Figure~\ref{fig:3}a,b shows $R$ at fixed $I$ (top panels) as well as its dependence on $I$ (bottom panels) as a function of $\nj$, while $\nleads$ is close to $-\ns/2$ (a) or to $+\ns/2$ (b), respectively.
As the quasiparticle current and the capacitance between the contacts influences the maximum supercurrent~\cite{tinkham2004}, we refer to this as the \lq{switching current}\rq $\Is$ from now on.
Comparing the measurements to the bulk behaviour shown in Fig.~\ref{fig:2}b, we still observe a finite $\Is$ when $\nj$ is tuned out of the superconducting state. That is, we observe the DC Josephson effect.
Note that when $\nj$=$\nleads$ the entire device is superconducting (shaded regions in Figs.~\ref{fig:3}c,d).
In Fig.~\ref{fig:3}a,b the JJ length $\Lj=\SI{100}{nm}$ is on the order of the estimated $\xi= \SI{70}{nm}$, indicative of the short JJ regime~\cite{tinkham2004,Li2016}.
Also, for the JJ with $\nleads=\SI{-1.5e12}{cm^{-2}}$ we observe signatures of ballistic transport~\cite{Calado2015} (SI Sec.~\ref{SI:FP}); $\Is$ is thus expected to be limited by the interfaces.
In addition to the JJs with $\Lj=\SI{100}{nm}$, our device allows us to realize other lengths, by using combinations of gates L (\SI{650}{nm}), R (\SI{650}{nm}) and C (\SI{100}{nm}).
Overall, $\Is$ as a function of $\nj$ is reduced for increased $\Lj$, as presented in Fig.~\ref{fig:3}c,d, and in the inset of of Fig.~\ref{fig:3}c, consistent with $\Lj\gg\xi$ for the $\Lj=\SI{650}{nm}$ and $\SI{1400}{nm}$ JJs, which means that these junctions are in the long-JJ regime~\cite{tinkham2004,Li2016} where the transport in the junction area is limiting $\Is$.

By changing $\nj$ we tune the JJs through the energy bands depicted in Fig.~\ref{fig:2}a, as reflected in the $R$ trace in Fig.~\ref{fig:3}a,b (highlighted with corresponding background colours).
We now analyse how $\Is$ changes with $\nj$, discuss this in terms of the junction area (N area) and the interfaces between the leads and the junction (N--S interfaces).
We observe step-like reductions in $\Is$ when $\nj$ crosses over to another band and a more gradual change of $\Is$ within a given band (Fig.~\ref{fig:3}a,b). 
These steps in $\Is$ occur when the Fermi level crosses a band gap at the CNP and at $\nj=\pm\ns$, and a correlated insulator gap at $n=\pm\ns/2$, which reduces the transparency of the N--S interfaces. 
We note that the abrupt changes can also have a contribution from differences between the eigenstates on each side of the interface, such as topological indices~\cite{nuckolls2020} or valley/spin flavours~\cite{Xie2020}.
Schottky barriers and interfacial defects can be neglected as the S--N interfaces are created in a single crystal.
Regarding the gradual change of $\Is$ within a band, we suspect the Fermi velocity $\vf$ to be most important for the proximity effect.
BCS theory predicts $\xi$ in the junction area to scale linearly with $\vf$ of the carriers in the N area~\cite{tinkham2004}.
Also, the N--S interface transparency is affected by a mismatch between $\vf$ in the leads and in the N area~\cite{Mortensen1999}, respectively.
The dramatic changes in $\vf$ with $n$ in the bands of MATBG~\cite{Suarez2010,Bistritzer2011,Nam2017,Guinea2019} therefore impacts the induced superconductivity. 
Note that $\Lj$ slightly changes due to the gate geometry used. For increasing $|\nj-\nleads|$, $\Lj$ could increase by approximately $\SI{20}{nm}$ (SI Sec.~\ref{SI:device}).

In the 1$^{\text{st}}$ moir\'{e} bands (blue and green) we observe rather large $\Is$ values, in line with the weakly insulating states at $\pm\ns/2$ and the CNP, leading to a significant $\Is$ even if the junction is $\SI{1400}{nm}$ long (Fig.~\ref{fig:3}c).
For $\Lj=\SI{650}{nm}$ we observe a non-monotonic behaviour of $\Is$ between $-\ns/2$ and the CNP (red arrow in Fig.~\ref{fig:3}c)). This could be due to an optimum in $\vf$, or might be caused by electron correlations that are already present, yet not strong enough to cause a superconducting ground state in our device~\cite{Lu2019}.

An interesting JJ is realized once $\nj$ is tuned to the opposite superconducting dome. 
The switching current $\Is$ is increased around $-\ns/2$ for all three $\Lj$ values as shown in Fig.~\ref{fig:3}b,d, as well as for $\Lj=\SI{650}{nm}$ and $\SI{1400}{nm}$ as shown in Fig.~\ref{fig:3}c.
If the leads and junction in this S--S--S configuration exhibit the same superconducting state, $\Is$ would solely depend on the interfaces and not on $\Lj$.
We observe, however, a dependence on $\Lj$ in both Fig.~\ref{fig:3}c and d, suggesting that the superconductivity in the two domes differ.
Such a scenario is an interesting avenue for future research, as it could be possible that we formed a JJ with two distinct superconductors.

When we tune $\nj$ into the 2$^{\text{nd}}$ and 3$^{\text{rd}}$ moir\'{e} band (violet colours in Fig.~\ref{fig:2}a), $\Is$ in Fig.~\ref{fig:3}(a,b) is greatly reduced, due to a crossing of the gap ($\pm\ns$) at the N--S interfaces.
In addition, another step in $R$ is observed for both the valence bands in Fig.~\ref{fig:3}a, as well as for the conduction bands in Fig.~\ref{fig:3}b.
At the same time, $\Is$ is either decreased or even reduced to zero, respectively.
Based on this observation and supported by Fabry-Pérot oscillations (SI Sec.~\ref{SI:FP}), we infer that we observe the band edge of the 3$^{\text{rd}}$ moir\'{e} band, and this band is therefore offset in energy with respect to the 2$^{\text{nd}}$ moir\'{e} band (Fig.~\ref{fig:2}a).
While before we connected the abrupt changes of $R$ and $\Is$ to the Fermi level crossing a gap, this is unlikely to be the case here, as we do not observe a resistance peak in Fig.\ref{fig:2}a around $n=\SI{-4.6e12}{cm^{-2}}$.
We speculate that the transition could occur because of a sudden change in $\vf$ (mismatch) or a mismatch in eigenstates in the leads and the junction.

\begin{figure}
\centering
\includegraphics[width=\textwidth]{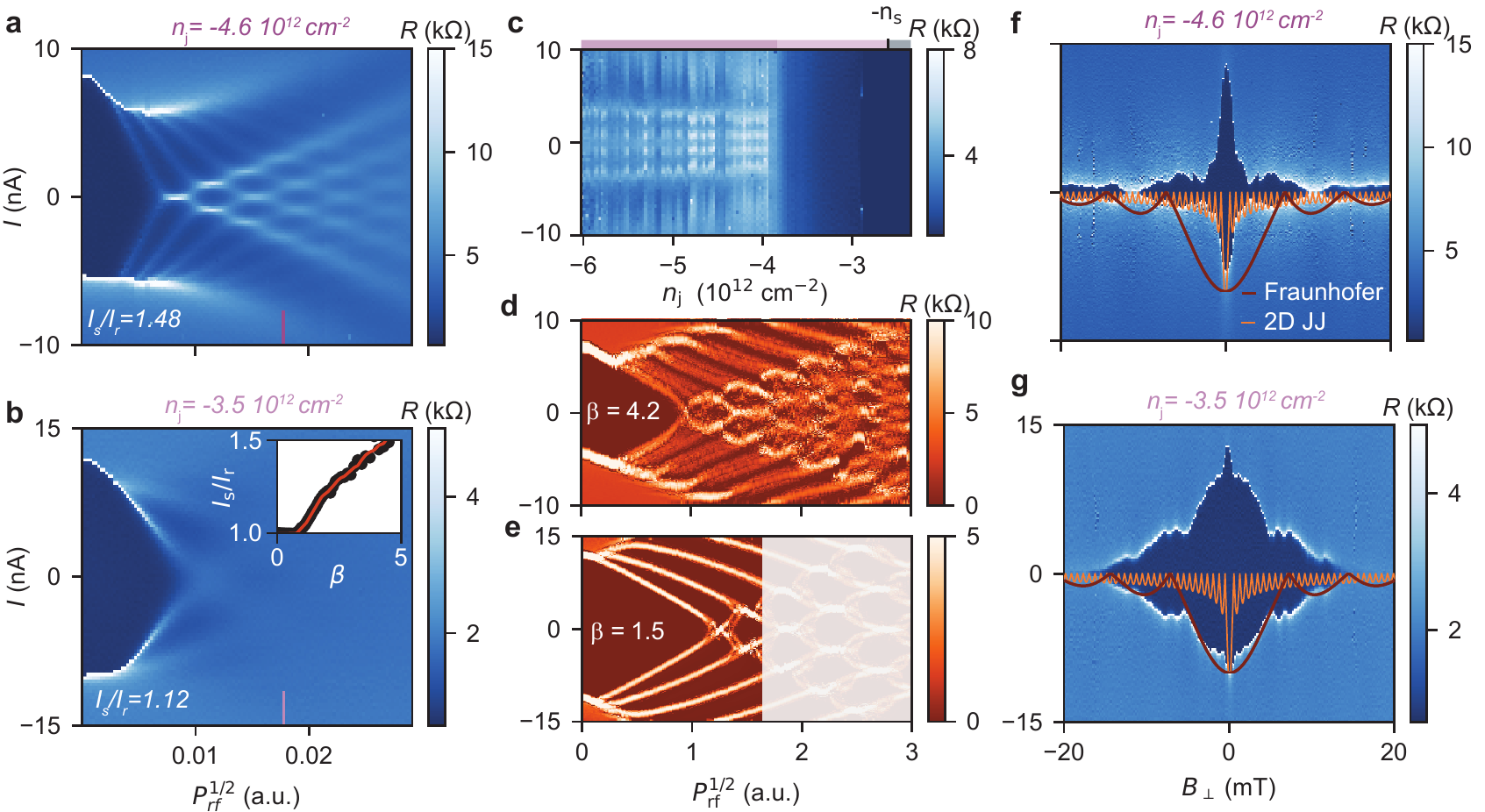}
\caption{\textbf{AC Josephson effects for the 2$^{\text{nd}}$ and 3$^{\text{rd}}$ moir\'{e} band.}
	\textbf{a,b}, Measurement of the Shapiro steps, visible in $R$ versus $I$ for different $P^{1/2}_\mathrm{rf}$, for $\nj$ as indicated. The hysteresis $\Is/\Ir$ is given. The inset in \textbf{b} shows the calculated $\Is/\Ir$ for different $\beta$.
	\textbf{c}, $R$ versus $I$ measured as a function of $\nj$ at $P^{1/2}_\mathrm{rf}$ indicated in \textbf{c,d}. The colour bar on top refers to the bands in Fig.~\ref{fig:2}a.
	\textbf{d,e}, Shapiro steps simulated using the resistively and capacitively shunted junction (RCSJ) model (see Methods) using the $\beta$ value indicated, $R_\text{j} = \SI{2.5}{k\Omega}$ and $R_\text{j} = \SI{1.14}{k\Omega}$ for the upper and lower panels, respectively, and $C_\textbf{j} = \SI{32}{fF}$ for both.
	The shaded area in \textbf{e} is added for better comparability.
	\textbf{f,g}, $R$ as a function of $I$ and $B$ shows the magnetic interference pattern of the $\SI{100}{nm}$ JJ with $\nleads=\SI{-1.5e12}{cm^{-2}}$ for the $\nj$ value indicated. For both scenarios we plotted a theoretical model for a 3D JJ and for a JJ in a 2D superconductor.
}\label{fig:4}
\end{figure}

We further studied this $\nj$-regime for the $\SI{100}{nm}$ JJ with $\nleads=\SI{-1.5e12}{cm^{-2}}$ by measuring the AC Josephson effect.
As $\nj$ is tuned into the 2$^{\text{nd}}$ and 3$^{\text{rd}}$ moir\'{e} band, where the N--S interfaces are less transparent, we observe the AC Josephson effect in the form of Shapiro steps~\cite{Shapiro1963}.
In the 3$^{\text{rd}}$ moir\'{e} band ($\nj$=\SI{-4.6e12}{cm^{-2}}) the steps have the trivial height of $\Delta V=(h/2e)\times f_\text{rf}$ for both frequencies of $f_\text{rf}=\SI{5}{Ghz}$ and $\SI{1.81}{Ghz}$, see Fig.~\ref{fig:1}c,d, and Fig.~\ref{fig:4}a, respectively, and SI Sec.~\ref{SI:Shapiro}.
Also, no integer steps are missing or only occurring at a specific $P^{1/2}_\mathrm{rf}$, as could be expected for a topological JJ~\cite{Bocquillon2017}. Therefore, our data suggests that the superconductivity around $\ns/2$ is of trivial rather than of topological nature.

The Shapiro-step spectrum for $\nj$ tuned in the 2$^{\text{nd}}$ moir\'{e} band only ($\nj=\SI{-3.5e12}{cm^{-2}}$) in Fig.~\ref{fig:4}b shows a dependence on $P^{1/2}_\text{rf}$ very different from that in Fig.~\ref{fig:4}a.
The steps are less dense and fade more quickly with respect to $P^{1/2}_\text{rf}$. 
We attribute the fading to thermal and impurity smearing of $R$, impacting the lower-valued $R$-peaks in Fig.~\ref{fig:4}b more than in Fig.~\ref{fig:4}a.
In addition, the hysteresis ($\Is/\Ir$) differs, suggesting a change in the JJ that can be traced back to the onset of the 3$^\text{rd}$ moir\'{e} band as the $\nj$-dependence in Fig.~\ref{fig:4}e reveals.

We compare the trend of this JJ behaviour to simulations using the resistively and capacitively shunted junction model~\cite{tinkham2004}(see Methods for details).
In this model, the resistance $R_\text{j}$, allowing for the flow of normal current, and $C_\text{j}$, the capacitance between the leads, are both connected in parallel to an ideal JJ with a given $\Ic$. 
Together, these parameters are captured in the Stewart--McCumber parameter $\beta\propto\Ic R_\text{j}^2 C_\text{j}$. 
In the inset of Fig.~\ref{fig:4}b, we show the simulated $\Is/\Ir$ as a function of $\beta$, and we estimate $\beta = 4.2$ and $\beta = 1.5$ for the $\Is/\Ir$ from Fig.~\ref{fig:4}a,b, respectively.
Then, we simulate Shapiro-step spectra using $\beta$, the measured $\Is$, a constant $C_\text{j}$ (the device geometry is constant), and 
we choose $R_\text{j}=\SI{2.5}{k\Omega}$ based on the normal-state resistance for $\beta = 4.2$, of which $R_\text{j}=\SI{1.14}{k\Omega}$ for $\beta = 1.5$ follows from the earlier-mentioned constraint on $C_\text{j}$.
The simulated spectra in Fig.~\ref{fig:4}d qualitatively reproduce the measured spectra of Fig.~\ref{fig:4}a,b.
As the smearing is not captured in the model, we added a semi-transparent box in Fig.~\ref{fig:4}e for better comparability.
The increase of $R_\text{j}$ when entering the $\mathrm{3^{rd}}$ moiré band means that the quasiparticle transport through the junction is reduced, which can be important for future device applications such as detectors~\cite{tinkham2004,Braginski2019}.

Lastly, we report magnetic interference measurements in this $\nj$-regime and compare them to theoretical models (see Methods).
In a textbook 3D JJ, applying $\Bp$ results in a modulation of $\Is$ described by a Fraunhofer pattern with the periodicity $\Delta\Bp=h/2e/(\Lj W)$, and $\Is$-maxima that scale with $1/\Bp$~\cite{tinkham2004}.
For a JJ in a 2D superconductor, $\Bp$ also penetrates the leads, and $\Delta\Bp=1.8\Phi_0/W^2$ and the $\Is$-maxima scale with $1/\Bp^{1/2}$~\cite{Moshe2008}.
In Fig.~\ref{fig:4}f at $\nj=\SI{-4.6e12}{cm^{-2}}$, the $\Is$-maxima follow the $1/B^{1/2}$ behaviour, but the periodicity matches neither theoretical pattern.
The ill-defined periodicity is likely to be caused by disorder in the junction area and the leads due to twist-angle inhomogeneity~\cite{Uri2020}.
For the interference pattern shown in Fig.~\ref{fig:4}g, measured at $\nj$=\SI{-3.48e12}{cm^{-2}}, $\Is$ is closer to the $1/\Bp$ scaling of a Fraunhofer pattern. 
However, it also lacks a clear periodicity, even though the smaller $\Is$ oscillations on the larger background compare better to the 2D JJ pattern.
Although we find clearly different interference patterns, further theoretical and experimental work is needed to understand the patterns themselves.

In conclusion, we have shown a proof-of-principle gate-defined superconducting device in MATBG, and explored this JJ by tuning through the different moir\'{e} bands. We defined JJs with different lengths and different superconducting states in a single device, establishing the flexibility of our approach. This work enables future technological applications, by virtue of its planar and gate-tunable superconductivity~\cite{Wendin2017,Liu2019}, as well as future investigations of the nature of the superconducting state in MATBG, for example by fabricating superconducting quantum interference devices or by creating JJs with a ferromagnetic barrier.

\section{Methods}
\subsection{Device Fabrication}
We assemble the heterostructure using the dry pick-up method~\cite{Kim2016}. As a first step, a large ($>\SI{40}{\micro m}\times\SI{40}{\micro m}$) mechanically exfoliated graphene flake is identified on a p:Si/SiO$_2$ ($\SI{285}{nm}$) wafer. The graphene flake is scratched into two pieces using a sharp ($\SI{2.5}{\micro m}$ tip diameter) tungsten needle which is controlled by a micromanipulator. Using a PDMS/PC stamp, the top hBN layer ($\SI{27}{nm}$) is picked up at a $T=\SI{80}{\degreeCelsius}$. At $\SI{40}{\degreeCelsius}$, we pick up the first graphene piece, rotate the stage by $\SI{1.1}{\degree}$ and pick up the second piece. The twisted graphene is encapsulated with a bottom hBN flake ($\SI{24}{nm}$), the flake is in contact with the stack at $\SI{40}{\degreeCelsius}$ and picked up at $\SI{80}{\degreeCelsius}$. At the same temperature, the graphite layer is picked up. PC is detached from PDMS at $\SI{150}{\degreeCelsius}$ and removed in dichloromethane. The top and bottom hBN thicknesses are determined using atomic force microscopy (AFM). AFM images serve to identify clean and bubble-free regions of the stack. Next, we define 1D contacts~\cite{Wang2013} to graphene by electron beam lithography (EBL), reactive ion etching (RIE, CHF$_3$/O$_2$, $40/4$sccm, $\SI{60}{W}$, with a $\SI{0.6}{nm/s}$ hBN etching rate, for $\SI{68}{s}$) and electron beam evaporation (Cr/Au, $10/\SI{70}{nm}$). Then, top-gates are defined by EBL (Cr/Au, $10/\SI{60}{nm}$) and the graphene is etched (RIE, CHF$_3$/O$_2$, $40/4$sccm, $\SI{60}{W}$, $\SI{120}{s}$, followed by O$_2$, $40$sccm, $\SI{60}{W}$, $\SI{10}{s}$). Using atomic layer deposition (ALD), $\SI{30}{nm}$ of AlO$_x$ are deposited at $T=\SI{150}{\degreeCelsius}$. In a final step, we define the central gates by EBL and deposit Cr/Au, $10/\SI{110}{nm}$.

\subsection{Measurement setup}
We use a DC voltage source built in-house with a $\SI{10}{M\Omega}$ connected in series to generate the bias current, amplify the voltage using a voltage amplifier built in-house and measure it with an Hewlett Packard 34401A multimeter. We measure in a 2 terminal setup, and subtracted a series resistance of $\SI{785}{\Omega}$ in Fig.~\ref{fig:1}b. The bottom gate, L gate and R gate are each connected to a DC voltage source. The C gate is connected to a DC voltage source and, using a bias T with $R=\SI{10}{k\Omega}$ and $C=\SI{10}{nF}$, to a Rohde und Schwarz SMB 100A signal generator. 

\subsection{RCSJ model}
We model our JJ with the resistively and capacitively shunted junction (RCSJ) model \cite{tinkham2004}. 
Replacing the voltage across the junction, $V$, with the superconducting phase between the two electrodes of the junction, $\phi$, using $V = (\Phi_0/2\pi)\times d\phi/dt$ \cite{Josephson1962}, we obtain the second order differential equation with reduced units \cite{Orlando1991}.

\begin{equation}
\label{eq:RCSJ_1}
    \frac{I}{I_\text{c}} = \sin{\phi} + \frac{d\phi}{d\tilde{\tau}} + \beta \frac{d^2 \phi}{d^2 \tilde{\tau}^2},
\end{equation}

containing the Stewart--McCumber parameter

\begin{equation}
    \beta = \frac{2\pi R_\text{j}^2 C_\text{j} \Ic}{\Phi_0},
\end{equation}

and $\tilde{\tau} = t/\tau_\text{J}$ with

\begin{equation}
    \tau_\text{J} = \frac{\Phi_0}{2\pi R_\text{j} \Ic}
\end{equation}

being the Josephson time constant. \newline

The resistance $R_\text{j}$ depends on the state of the junction. 
When operating it beyond its critical current, $R_\text{j}$ corresponds to $R_\text{n}$, the normal state resistance of the junction. 
When a supercurrent is flowing, $R_\text{j}$ accounts for tunnelling of quasiparticles through the junction. 
The quasiparticle resistance $R_\text{qp}$ is, in general, expected to be higher than $R_\text{n}$. 
It is hard to estimate the value of $R_\text{qp}$ and its dependence on the current $I$, espacially since our knowledge about the electronic state in the junction area is limited.
We have therefore decided to assume $R_\text{j}$ is on the order of $R_\text{n}$.

The switching current to retrapping current ratio $\Is/I_\text{r}$ is directly related to $\beta$~\cite{tinkham2004}. 
Since there exists no exact analytical relation between the parameters, we compute it (inset in Fig. \ref{fig:4}b). We then obtain $\beta_\text{C} = 4.2$ and $\beta_\text{C} = 1.5$ for $\nj = -4.6 \times 10^{12} \mathrm{cm^{-2}}$ and $\nj = -3.5 \times 10^{12} \mathrm{cm^{-2}}$ respectively. \newline
    
For modelling the behavior of an RF-driven Josephson junction we add a sinusoidal term to the current bias term of equation (\ref{eq:RCSJ_1}), thus obtaining

\begin{equation}
\label{eq:RCSJ_2}
    \frac{I}{I_\text{c}} + \frac{I_\text{RF}}{I_\text{c}}\sin{\tilde{\omega}\tilde{\tau}} = \sin{\phi} + \frac{d\phi}{d\tilde{\tau}} + \beta_\text{C} \frac{d^2 \phi}{d^2 \tilde{\tau}^2}
\end{equation}

with $\tilde{\omega} = \omega_\text{RF} \tau_\text{J}$ the reduced angular frequency of the RF-source~\cite{Russer1972}.
    
In this mathematical treatment we neglect the frequency-dependent coupling of the RF-signal to the junction. 
Furthermore, we assume the relationship between the sinusoidal voltage of the RF-wave and the resulting AC current through the junction to be linear. 

The reduced angular frequency, $\tilde{\omega}$, is the relevant parameter for the observation of the Shapiro steps. 
Only if the condition $\tilde{\omega} \ll 1$ is fulfilled, the phase of the junction is properly locked to that of the RF-source, and Shapiro steps can be observed~\cite{LeCalvez2019}. 
Three parameters determine $\tilde{\omega}$: The RF-frequency, which is in our simulations always fixed to $\omega_\text{RF} = \SI{1.81}{GHz}$, the critical current of the junction, which we fix to either $I_\text{C} = \SI{7}{nA}$ or $I_\text{C} = \SI{12}{nA}$, and the resistance $R$ of the junction. 
As we mentioned previously, we extract experimentally $\beta_\text{C} \propto I_\text{C} C R^2$, which leaves us with two free parameters for our modelling, $R$ and $C$. 
We start at $\beta_\text{C} = 4.2$ and choose $R = \SI{2.5}{k\Omega}$ since it is of the order of the normal state resistance for $\nj = \SI{-4.6e12}{cm^{-2}}$. 
We compute the resulting capacitance $C = 31.2 \mathrm{fF}$ and perform the simulation (\ref{fig:4}d). 
When moving to $\beta_\text{C} = 1.5$, for simulating the $\nj = \SI{-3.5e12}{cm^{-2}}$ case, we keep $C$ constant and obtain $R = \SI{1.14}{k\Omega}$, obtaining the result shown in Fig. \ref{fig:4}e.

\subsection{Magnetic interference patterns}
Here we provide the equations used to plot the theoretical magnetic interference patterns.
The Fraunhofer interference pattern follows
\begin{equation}
    I(\Bp)=\Ic\left|\frac{\sin(\pi \Bp \Lj W/\Phi_0)}{\pi \Bp \Lj W/\Phi_0}\right|.
\end{equation}
For the Josephson junction in a 2D superconductor we use the approximation
\begin{equation}
    I(\Bp)=0.61 \frac{\Ic}{W} \sqrt{\frac{\Phi_0}{|\Bp|}} \left|\cos(1.72|\Bp|W^2/\Phi_0-\pi/4)\right|.
\end{equation}
given in Ref.~\citenum{Moshe2008}.

\newpage
\bibliographystyle{naturemag}
\bibliography{JJ_MATBG_arxiv1.bbl}

\section{Acknowledgements}
We thank Benedikt Kratochwil, Peter Maerki and the staff of the ETH cleanroom facility FIRST for technical support, and Tobias Wolf for usefull discussions.
We acknowledges support from the Graphene Flagship and from the European Union's Horizon 2020 research and innovation programme under grant agreement No 862660/QUANTUM E LEAPS.
K.W. and T.T. acknowledge support from the Elemental Strategy Initiative conducted by the MEXT, Japan ,Grant Number JPMXP0112101001,  JSPS KAKENHI Grant Number JP20H00354 and the CREST(JPMJCR15F3), JST.
E.P acknowledges support from La Caixa Foundation.

\section{Author information}
\subsection{Corresponding authors}
Correspondence and requests for materials should be addressed to Folkert K. de Vries (devriesf@phys.ethz.ch) and Peter Rickhaus (peterri@phys.ethz.ch).

\subsection{Author contributions}
E.P. made the stack and P.R. fabricated the device, with assistance from G.Z.
T.T. and K.W. supplied the hBN crystals.
F.K.d.V and E.P. performed the measurements.
E.P. performed the simulations.
T.I., K.E. and P.R. supervised the project.
F.K.d.V., P.R. and E.P. wrote the manuscript with comments from all authors.

\subsection{Competing interests}
The authors declare no competing financial interests.

\clearpage
\begin{center}
    \Large{\textbf{Supplementary Information}}
\end{center}
\setcounter{figure}{0}
\setcounter{page}{1}
\renewcommand{\thefigure}{S\arabic{figure}}
\renewcommand{\theequation}{S\arabic{equation}}
\setcounter{secnumdepth}{2}

\section{Device}\label{SI:device}
\begin{figure}[h]
\centering
\includegraphics[width=1\textwidth]{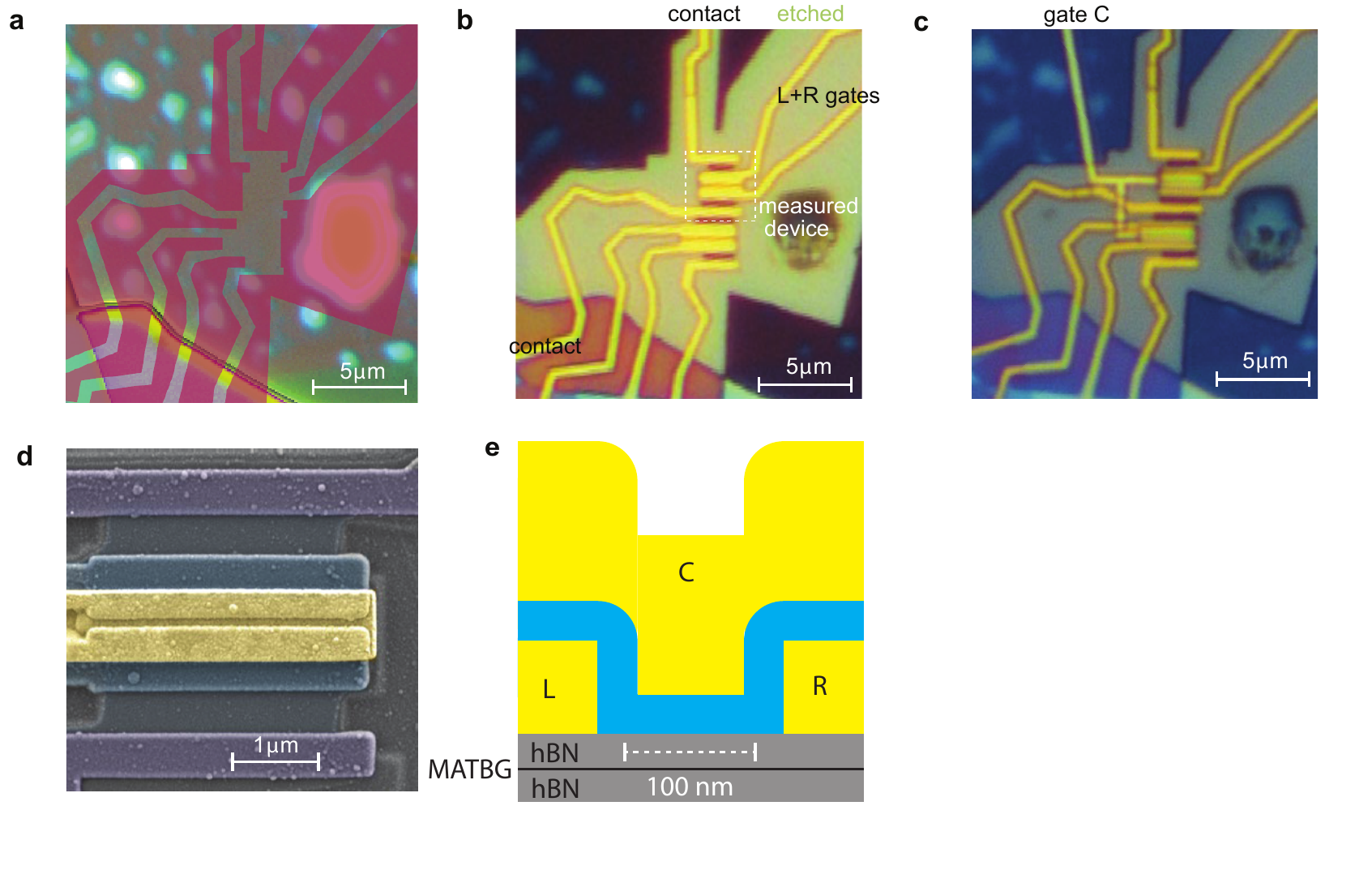}
\caption{\textbf{Fabrication images of the device.}
\textbf{a} Optical image before fabrication, the red shaded area is the etch mask.
\textbf{b} Optical image before and \textbf{c} after deposition of the central gate electrodes.
\textbf{d} Scanning electron microscopy image of a similar device (not titled). 
\textbf{e} Schematic cross-section of the device, all layers are to scale.
}\label{suppfig:1}
\end{figure}

In Fig.~\ref{suppfig:1}a-c we show optical images of the device fabrication process. The measured device is marked with a dashed square. Figure~\ref{suppfig:1}a shows that the device is placed carefully in-between the bubbles in the MATBG. The bubbles are expected to lead to twist angle inhomogeneity~\cite{Uri2020}. In transport, we find that the lower device has a larger twist angle ($\theta=1.5^\circ$) and does not show superconductivity. The narrow gap between the L+R gates is well resolved in the SEM image in Fig.~\ref{suppfig:1}c, allowing us to identify a typical length of $\SI{100}{nm}$. The SEM image is taken from a similar device fabricated in the same batch and with similar geometry as the measured device (i.e. same length of L+R gates and same gap between these gates in the design file).
The schematic cross-section in Fig.~\ref{suppfig:1}e shows this length. Also, we estimate that the size of the gated region below gate C can vary by $\pm\SI{10}{nm}$. 

\newpage
\section{Critical current dependence on temperature and magnetic field}\label{SI:IcTB}
\begin{figure}[h]
\centering
\includegraphics[width=1\textwidth]{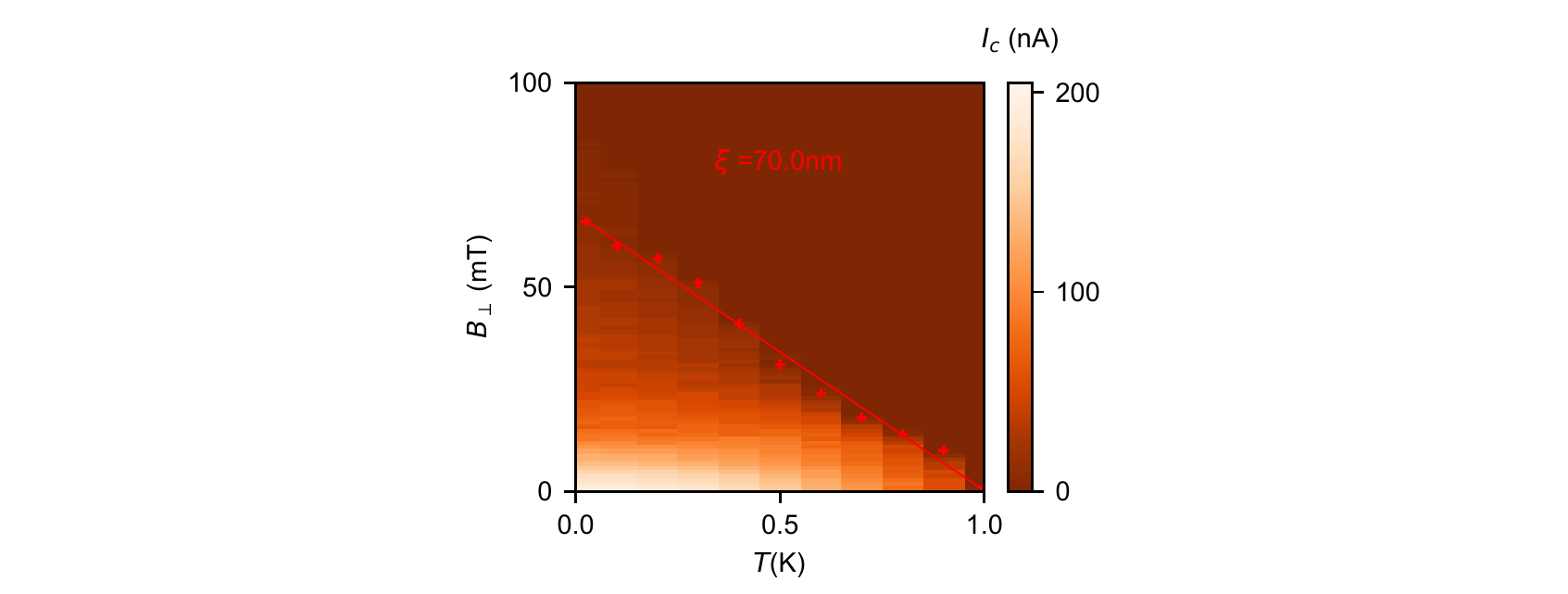}
\caption{$\Ic$  verus $\Bp$ and $T$. The critical magnetic field is extracted and plotted as red crosses. A fit with the Ginzburg--Landau theory provides us with the estimate of the superconducting coherence length $\xi$ indicated.
}\label{suppfig:2}
\end{figure}

We measure $\Ic$ as a function of temperature $T$ and perpendicular magnetic field $\Bp$ at $n = \SI{-1.53e12}{cm^{-2}}$, and present the result in in Fig.~\ref{suppfig:2}.
The extracted critical magnetic field allows us to fit the Ginzburg--Landau equation $B_{\perp,c}=(\hbar/2e)/ \xi^2(1-T/T_\text{c})$ for two-dimensional superconductors~\cite{tinkham2004}, and obtain an estimate of the superconducting coherence length $\xi$.

\newpage
\section{Fabry-Pérot oscillations}\label{SI:FP}
\begin{figure}[h]
\centering
\includegraphics[width=1\textwidth]{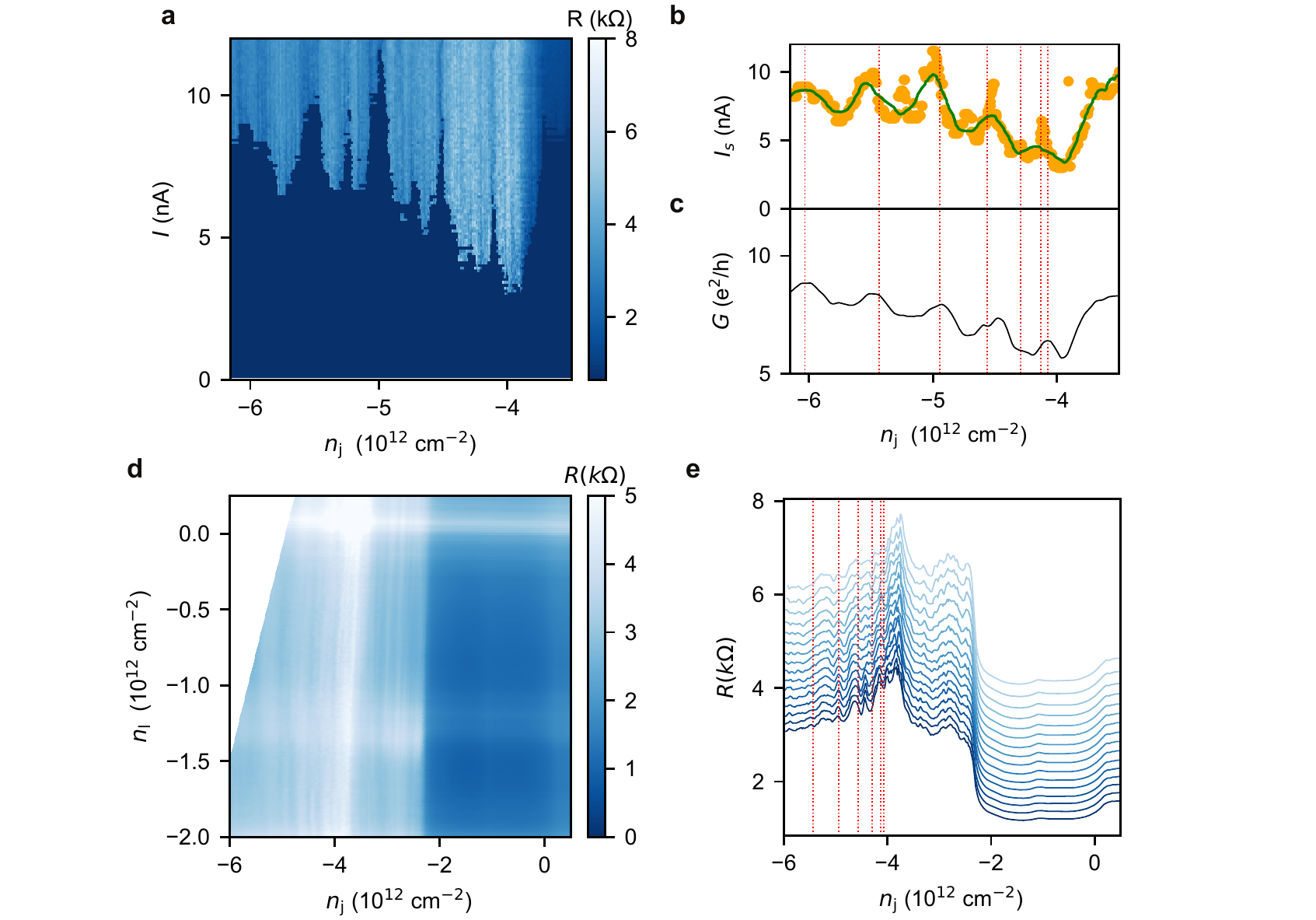}
\caption{\textbf{Fabry-Pérot oscillations at the onset of the 3$^\text{rd}$ moir\'{e} band.}
\textbf{a} Zoom in on $R$ versus $I$ and $\nj$ for the JJ of $\SI{100}{nm}$ and $\nleads=\SI{-1.5e12}{cm^{-2}}$.
\textbf{b} $\Is$ versus $\nj$ extracted from \textbf{a} and a smoothed curve fit through it.
\textbf{c} $G$ versus $\nj$ calculated at $I=\SI{350}{nA}$.
\textbf{d} $R$ as a function of $\nleads$ and $\nj$, at $I=\SI{350}{nA}$.
\textbf{e} Extracted line-traces from (a). The orange dotted lines mark conditions for destructive Fabry-Pérot interference, assuming a cavity length of $\SI{75}{nm}$.
}\label{suppfig:3}
\end{figure}

In addition to the step in $R$ observed in the dispersive bands, we investigate Fabry--Pérot resonances for the \SI{100}{nm} JJ with $\nleads=\SI{-1.5e12}{cm^{-2}}$, similar to earlier work in proximitized graphene JJs~\cite{Calado2015}.
In Fig.~\ref{suppfig:3}a,b we show a zoom-in to the region where $\Is$ is modulated by $\sqrt{\nj}$ for $\nj<\ns$. The conductance $G$ in Fig.~\ref{suppfig:3}c, recorded at $I=\SI{350}{nA}$, behaves in a similar fashion. 
Local maximum values of $\Ic$ and $G$ occur if the Fabry-Pérot condition is fulfilled. 
For constructive interference the condition reads $i\lambda = 2l$, where $i$ is an integer, $\lambda$ the charge carrier wavelength and $l$ the cavity length. The condition can be expressed as a function of density by using $\lambda=2\pi/k$ and $k=\sqrt{(4/g) n_i\pi}$, giving $n_i=\pm \pi i^2/l^2(4/g)$. The third valence band is expected to be 4-fold degenerate thus $g=4$.
The observed maxima in $\Ic$ and $G$ (Fig.~\ref{suppfig:3}b,c) are well captured by setting $l=\SI{75}{nm}$ (indicated by orange dotted lines) which is in good a agreement with $\Lj$. 

We further confirm the ballistic origin of the oscillations by investigating the dependence on $\nleads$, see Fig.~\ref{suppfig:3}d,e, where we plot $R(\nj,\nleads)$. The Fabry-Pérot oscillations show little dependence on $\nleads$, confirming that the oscillations originate from a region below gate C. Such behavior becomes clear in the line traces $R(\nj)$ for different $\nleads$, see Fig.~\ref{suppfig:3}e, where the $\nj$ at which we expect constructive Fabry-Pérot oscillations are indicated by orange dotted lines.

\newpage
\section{Shapiro step line traces}\label{SI:Shapiro}

\begin{figure}[h]
\centering
\includegraphics[width=1\textwidth]{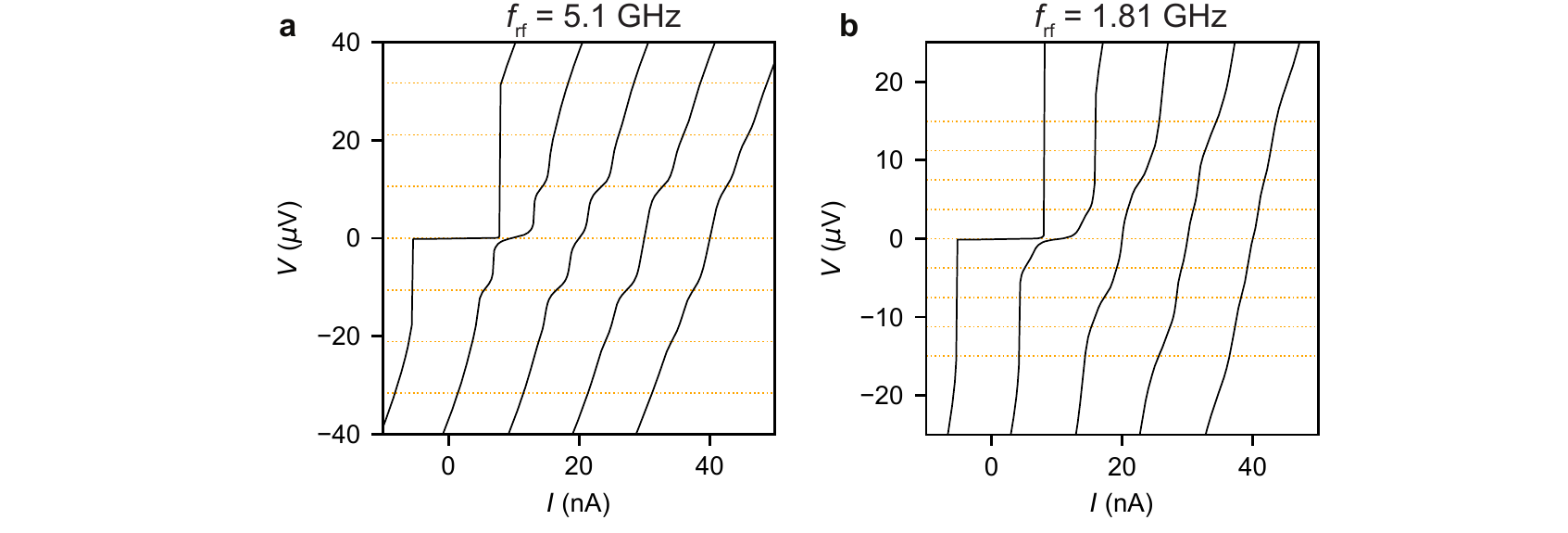}
\caption{\textbf{Line traces of Shapiro steps at different frequencies.}
\textbf{a,b} $V$ versus $I$ show shapiro step line traces for the JJ of $\SI{100}{nm}$ with $\nj=\SI{-4.6e12}{cm^{-2}}$ $\nleads=\SI{-1.5e12}{cm^{-2}}$ for the frequencies indicated. The traces are offset by $\SI{10}{nA}$ and taken at $P^{1/2}_\mathrm{rf}= 0, 0.015, 0.022, 0.028, 0.035$ for \textbf{a}, and $P^{1/2}_\mathrm{rf}= 0, 0.005, 0.009, 0.014, 0.019$. The orange dashed lines mark the expected steps at multiples of $\Delta V=h f_\text{rf}/ 2e$.
}\label{suppfig:4}
\end{figure}

In Fig.~\ref{suppfig:4}a,b we show line traces of the Shapiro-step spectra at both frequencies \SI{5.1}{GHz} and \SI{1.8}{GHz}. We observe that the steps occur at the expected voltage following $\Delta V=h f_\text{rf}/ 2e$.

\end{document}